\documentclass[conference]{IEEEtran}
\IEEEoverridecommandlockouts
\usepackage{amsmath,amssymb,amsfonts}
\usepackage{algorithmic}
\usepackage{graphicx}
\usepackage{textcomp}
\usepackage{xcolor}
\def\BibTeX{{\rm B\kern-.05em{\sc i\kern-.025em b}\kern-.08em
    T\kern-.1667em\lower.7ex\hbox{E}\kern-.125emX}}

\usepackage{textcmds}

\usepackage{booktabs}
\usepackage{tabularx}
\usepackage{makecell}
\usepackage{tabularray}

\usepackage[normalem]{ulem}

\usepackage{subcaption}

\usepackage{enumerate}
\usepackage[numbers,sort&compress]{natbib}

\usepackage[frozencache,cachedir=.]{minted} 
\newsavebox{\largestimage}
\newsavebox{\largestimagedd}
\newsavebox{\largestimageind}

\usepackage{soul}
\usepackage{tikz}
\usetikzlibrary{calc}
\usepackage{hyperref}
\usepackage{cleveref}

\makeatletter
\newif\if@anonymize

\@anonymizefalse  %

\if@anonymize
  \newcommand{\highlight@DoHighlight}{
    \fill [outer sep = -15pt, inner sep = 0pt, color=black]
          ($(begin highlight)+(0,8pt)$) rectangle ($(end highlight)+(0,-3pt)$) ;
  }

  \newcommand{\highlight@BeginHighlight}{
    \coordinate (begin highlight) at (0,0) ;
  }

  \newcommand{\highlight@EndHighlight}{
    \coordinate (end highlight) at (0,0) ;
  }

  \newdimen\highlight@previous
  \newdimen\highlight@current
  \newlength{\item@width}

  \DeclareRobustCommand*\anonymize{%
    \SOUL@setup
    \def\SOUL@preamble{%
      \begin{tikzpicture}[overlay, remember picture]
        \highlight@BeginHighlight
        \highlight@EndHighlight
      \end{tikzpicture}%
    }%
    \def\SOUL@postamble{%
      \begin{tikzpicture}[overlay, remember picture]
        \highlight@EndHighlight
        \highlight@DoHighlight
      \end{tikzpicture}%
    }%
    \def\SOUL@everyhyphen{%
      \discretionary{%
        \SOUL@setkern\SOUL@hyphkern
        \SOUL@sethyphenchar
        \tikz[overlay, remember picture] \highlight@EndHighlight ;%
      }{%
      }{%
        \SOUL@setkern\SOUL@charkern
      }%
    }%
    \def\SOUL@everyexhyphen##1{%
      \SOUL@setkern\SOUL@hyphkern
      \settowidth{\item@width}{##1}%
      \makebox[\item@width]{}%
      \discretionary{%
        \tikz[overlay, remember picture] \highlight@EndHighlight ;%
      }{%
      }{%
        \SOUL@setkern\SOUL@charkern
      }%
    }%
    \def\SOUL@everysyllable{%
      \begin{tikzpicture}[overlay, remember picture]
        \path let \p0 = (begin highlight), \p1 = (0,0) in \pgfextra
          \global\highlight@previous=\y0
          \global\highlight@current =\y1
        \endpgfextra (0,0) ;
        \ifdim\highlight@current < \highlight@previous
          \highlight@DoHighlight
          \highlight@BeginHighlight
        \fi
      \end{tikzpicture}%
      \settowidth{\item@width}{\the\SOUL@syllable}%
      \makebox[\item@width]{}%
      \tikz[overlay, remember picture] \highlight@EndHighlight ;%
    }%
    \SOUL@
  }
\else
  \newcommand{\anonymize}[1]{#1}
\fi
\makeatother

\begin{document}

\title{Identifying Defect-Inducing Changes in Visual Code}

\author{\IEEEauthorblockN{\anonymize{Kalvin Eng}}
\IEEEauthorblockA{
\anonymize{\textit{Quality, Verification \& Standards}} \\
\anonymize{\textit{Electronic Arts}}\\
\anonymize{Edmonton, Canada} \\
\anonymize{kalvin.eng@\{ualberta.ca, ea.com\}}}
\and
\IEEEauthorblockN{\anonymize{Abram Hindle}}
\IEEEauthorblockA{
\anonymize{\textit{Department of Computing Science}} \\
\anonymize{\textit{University of Alberta}}\\
\anonymize{Edmonton, Canada} \\
\anonymize{abram.hindle@ualberta.ca}}
\and
\IEEEauthorblockN{\anonymize{Alexander Senchenko}}
\IEEEauthorblockA{
\anonymize{\textit{Quality, Verification \& Standards}} \\
\anonymize{\textit{Electronic Arts}}\\
\anonymize{Vancouver, Canada} \\
\anonymize{asenchenko@ea.com}}
}

\maketitle

\begin{abstract}
Defects, or bugs, often form during software development. Identifying the root cause of defects is essential to improve code quality, evaluate testing methods, and support defect prediction. Examples of defect-inducing changes can be found using the 
SZZ algorithm to trace the textual history of defect-fixing changes back to the defect-inducing changes that they fix in line-based code. The line-based approach of the SZZ method is ineffective for visual code that represents source code graphically rather than textually. In this paper we adapt SZZ for visual code and present the \textit{SZZ Visual Code} (SZZ-VC) algorithm, that finds changes in visual code based on the differences of graphical elements rather than differences of lines to detect defect-inducing changes. We validated the algorithm for an industry-made AAA video game and 20 music visual programming defects across 12 open source projects. Our results show that SZZ-VC is feasible for detecting defects in visual code for 3 different visual programming languages.
\end{abstract}

\begin{IEEEkeywords}
visual programming, visual code, bugs, defects, version control
\end{IEEEkeywords}

\section{Introduction}
In this paper, we seek to find the commits that induce defects in visual code. Defects, which are commonly referred to as bugs, occur when unintended flaws are introduced into software during the software development process. Determining the defect-inducing code helps to look for similar defective code at commit time that can lead to future problems and helps developers to avoid them. The SZZ algorithm~\cite{sliwerski2005changes} has been developed to detect changes that cause defects, after the defect has been fixed, by looking for changes to the fixing lines of code leveraging textual differencing. Therefore, the SZZ algorithm does not naturally work on visual source code.

Visual code, also known as block-based code, low code, no-code, or visual scripts, is the result of a paradigm in programming that relies on visually manipulating graphical elements on a 2D canvas to create programs~\cite{myers1990taxonomies}. The syntax of visual code is visual (often nodes and edges, or puzzle pieces). In comparison, textual code is traditionally created as text on a line-by-line basis. 

For example, Burlet \textit{et al.}~\cite{burlet2015empirical} describe popular visual music programming languages such as Pure Data~\cite{puredata} and its proprietary counterpart Max/MSP~\cite{max-msp}. These languages allow users to programmatically arrange rectangular objects on the screen and connect them with lines called patch cords to generate sound and respond to human-computer interaction devices. They describe the rectangular objects as functions that manipulate audio signals or other data structures and have inlets and outlets corresponding to parameters and return values, respectively. Burlet \textit{et al.}\ describe the visual code of Pure Data and Max/MSP being represented as a \textit{patch} in a file that contains a collection of connected objects that perform a musical function. It is common for computer music applications to consist of several interconnected patches. 

The use of visual programming has been very effective in the domains of video game development~\cite{unity-vs, unity-playmaker, unreal-blueprints, schenk2013scriptease}, sound development~\cite{max-msp, puredata}, IoT~\cite{nodered}, and coding education~\cite{resnick2009scratch, patton2019app, harvey2013snap, maclaurin2011design}. Despite the prevalence of visual code in a variety of domains, many tools in software engineering have not been ported to visual source code.

At \emph{EA} (Electronic Arts), video game development teams have increasingly adopted visual code as their primary approach for constructing video game features such as triggers on maps and game rules. This shift to visual code allows for broader team participation in the software development process and motivates the need to develop a methodology for detecting defect-inducing changes. Changes to visual code might not be accurately detected with the traditional textual SZZ. Thus, a visual code SZZ approach is needed so that visual defect-inducing changes and defective visual code can be gathered for building visual code defect prediction models~\cite{senchenko2022supernova}. To the best of our knowledge, no methodology has been developed for detecting defect-inducing changes in visual code.

In this work, we contribute \textit{SZZ Visual Code} (SZZ-VC), an  adaptation of the SZZ algorithm to visual code to identify defect-fixing and defect-inducing changes in visual code. SZZ-VC keeps track of nodes and edges, instead of source code lines. We manually evaluate its feasibility on 20 defects across 12 open source projects and 30 defects in a AAA video game (a big budget game from a large studio).

\section{Background - SZZ Algorithm}
\label{og-szz}

\begin{figure*}[htbp]
\centerline{\includegraphics[width=0.95\linewidth]{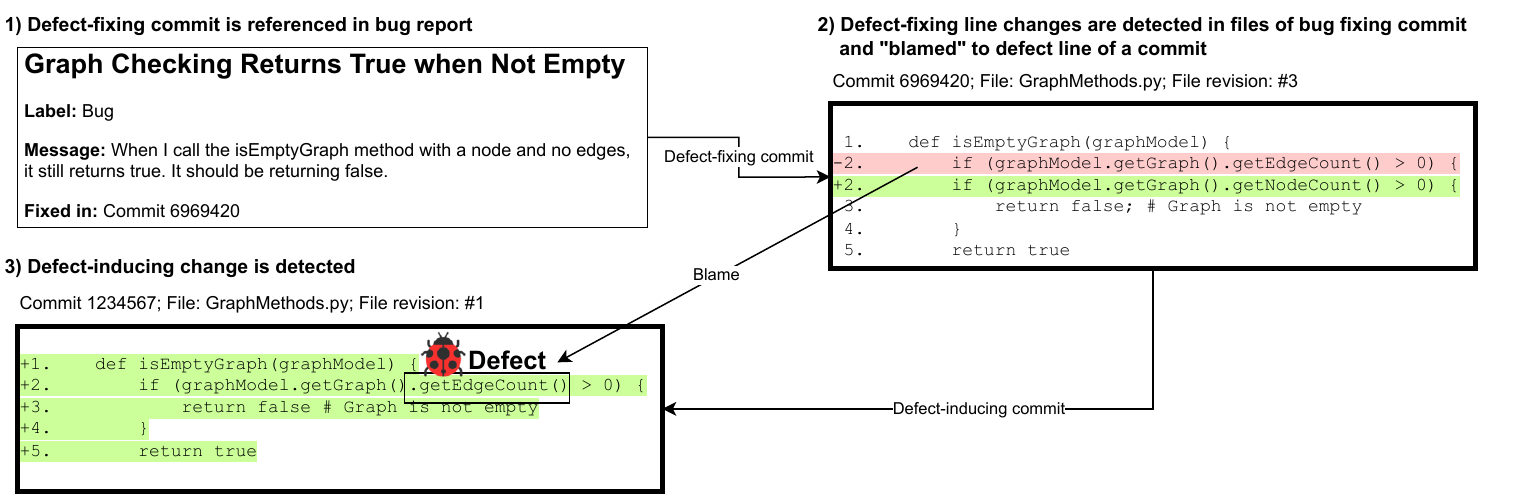}}
\caption{High-level example of SZZ algorithm to identify potential defect-inducing commits.}
\label{fig:szz_algo}
\end{figure*}

The SZZ algorithm is widely used in empirical software engineering to identify changes in textual code that may have caused defects~\cite{sliwerski2005changes}. It is often used to find defect-inducing targets for defect prediction models. \Cref{fig:szz_algo} illustrates how SZZ can detect potential defect-inducing changes from a bug report. The SZZ algorithm can be described in two parts.

In the first part, defect-fixing changes need to be found. Changes are represented as commits in a version control system, hence defect-fixing \qq{changes} can be interchangeably used with defect-fixing \qq{commits}. \citet{sliwerski2005changes} suggest that \textit{defect-fixing commits} can be found by looking for references in bug reports of issue trackers or by parsing commit messages that describe the commit as being a fix. 

In the second part, defect-inducing commits are found from defect-fixing commits. For each defect-fixing commit, each line of code that has been modified in the commit is backtracked through the history of the code within the version control repository to identify the \textit{previous commit} that originally introduced the line that was changed to fix the defect for all changed lines. The previous commit is a \textit{potential defect-inducing commit}, and a heuristic is run to filter out any unviable defect-inducing commits for a final set of filtered \textit{defect-inducing commits} that inject defects into code.

It should be noted that there are limitations in both parts of the original SZZ algorithm. Limitations in the first part include needing defect-fixing commits to be properly identified via commit messages or an issue tracker --- the quality of these defect-fixing annotations that accurately identify a commit as defect-fixing can widely differ~\cite{herzig2013s}. There can also be other sources of defect-fixing commits such as pull requests~\cite{bludau2022pr}. Sometimes, defect-fixing commits may not be actually be defect-fixing~\cite{herbold2022problems, neto2018impact}. Often, defect-fixing commits incorporate multiple files or changes that are not all fixes~\cite{da2016framework}. \citet{rodriguez2020bugs, rodriguez2018if} define defects as being intrinsic when defect-inducing changes can be found in the code, and extrinsic when defects are caused by external factors such as changes in API dependencies or changing requirements. Since extrinsic defects are caused externally of the code, they have no defect-inducing changes in the version control system and cannot be detected with the SZZ algorithm. 

In the second part of the algorithm, where defect-inducing changes are found, many changes can be false-positives. This can be due to changes being cosmetic such as modifying comments, blank lines, and formatting~\cite{kim2006automatic}. Code refactorings may also not be defect-inducing as they might just move defect-inducing code around~\cite{neto2018impact, neto2019revisiting}. As well, multiple code changes at once can mean that not all changes are defect-fixing leading to false-positives~\cite{da2016framework}. In addition to multiple changes, it is also difficult to identify defect-inducing changes if there are multiple files in the defect-fixing change as it is uncertain which files contribute to a fix~\cite{kim2006automatic}. False-positives can also arise from inaccurately tracing the historical textual differences in lines of code~\cite{williams2008szz, davies2014comparing, sinha2010buginnings, borg2019szz, rezk2021ghost, sahal2018identifying, bludau2022pr, rosa2021evaluating}. Moreover, the first detected change to a line of code that is fixed may not necessarily be the defect-inducing change~\cite{bao2022v}. Furthermore, fixes can be non-functional defects (e.g.\ security or performance issues like software running slow) or functional defects (i.e.\ impacting specific functionality like being unable to press a button). \citet{quach2021empirical} find that non-functional defects can hamper the detection of defect-inducing changes.

Several variants of the SZZ algorithm for textual code have been proposed~\cite{kim2006automatic, williams2008szz, davies2014comparing, neto2018impact, neto2019revisiting, borg2019szz, bao2022v, bludau2022pr, da2016framework, sahal2018identifying, sinha2010buginnings} and reviewed~\cite{neto2019revisiting,  herbold2022problems, rosa2021evaluating, da2016framework, rodriguez2020bugs, quach2021empirical, rosa2023comprehensive, rezk2021ghost, rodriguez2018if}. The most related work to visual code is the SZZ variant introduced by \citet{sinha2010buginnings} that performs graph-based differencing of \textit{Program Dependence Graphs}~\cite{ottenstein1984program} regions in textual code. In our work, we also perform tree-based differencing to identify changes in a tree-based intermediate representation of visual code.

To the best of our knowledge, none of the SZZ variants have accounted for changes in visual programming languages since changes in graphical elements such as nodes and edges are different from changes in lines of code and naturally cannot be compared line-by-line even if visual code is stored in a textual format like JSON.

\section{SZZ Visual Code (SZZ-VC)}
\begin{figure*}[htbp]
\centerline{\hspace{1cm}\includegraphics[width=0.90\linewidth]{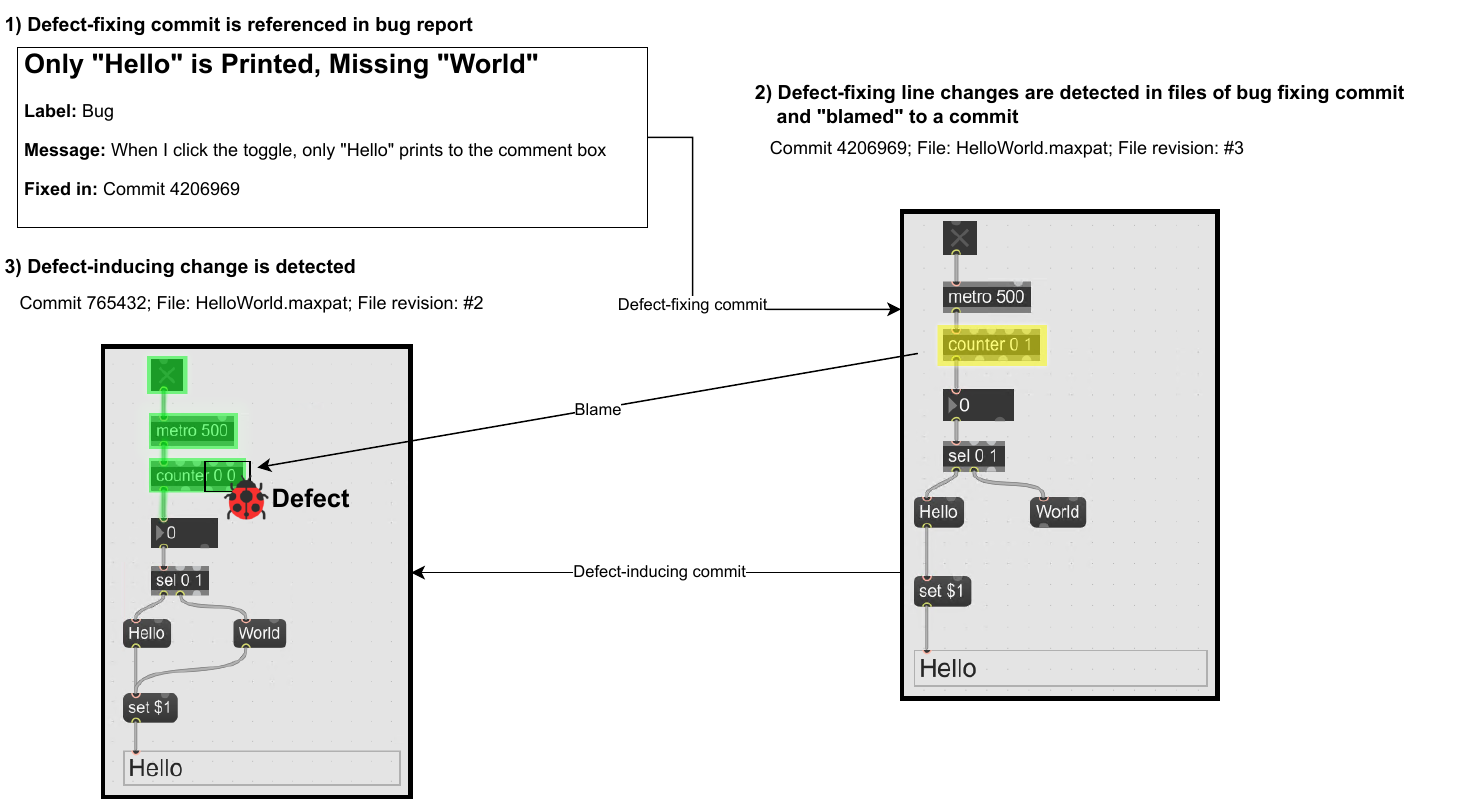}}
\caption{High-level example of SZZ-VC on a Max/MSP patch to detect potential defect-inducing commits from a defect-fixing commit. The fix is changing the counter max range from 0 to 1. Yellow indicates a modification, while green indicates an addition.}
\label{fig:szz_visual_algo}
\end{figure*}

\textit{SZZ Visual Code} (SZZ-VC) is adapted from the SZZ algorithm for node and edge based visual code. We present below our visual code adaptation for the first part of SZZ in steps (1) and (2) that identifies suitable defect-fixing commits, and for the second part of SZZ in steps (3) and (4) which seeks to find defect-inducing changes. Steps (1)-(3) can be visualized in \Cref{fig:szz_visual_algo}.  SZZ-VC works as follows:
\begingroup
\renewcommand\labelenumi{(\theenumi)}
\begin{enumerate}
    \item Defect-fixing commits are identified from either commit messages that link to a bug report in the issue tracker or bug reports in the issue tracker that link to a commit message similar to the first part of SZZ described in \Cref{og-szz}.
    \item For each of the defect-fixing commits, the files containing visual code are identified with heuristics such as choosing known visual code file extension types. If a visual code file is detected, then the defect-fixing commit is suitable to be used for finding defect-inducing commits.
    \item For each of the visual code files in the commit, the textual contents are serialized into an intermediate representation (IR) that represents the visual code's nodes and edges in a tree based format where nodes are at the top of the tree and contain edges. The subtrees of the IR are node contents in the visual code which is compared against previous IR versions of the visual code files to find changes in nodes and edges. \textbf{We assume that a file is \textit{potentially defect-inducing} for the file(s) in the defect-fixing file history that modify the node contents that were modified prior to the change that the defect-fixing commit fixes.} 
    \item Finally, a filtering step can be applied to remove any potential defect-inducing changes that are false positives (e.g.\ a heuristic to remove commits that have been committed after the time that a defect has been reported).
\end{enumerate}
\endgroup

\begin{figure*}[htbp]
\centering
\savebox{\largestimage}{\includegraphics[width=0.45\textwidth]{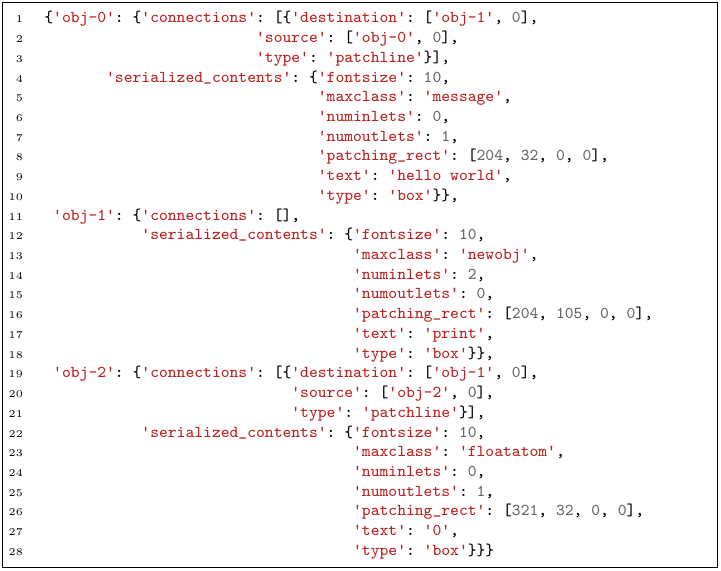}}%
\subcaptionbox[Short Subcaption]{%
    Pure Data visual code \qq{hello world} patch annotated with IR's root subtree id's (top) and textual code (bottom).%
    \label{subfig:ir_1}%
}
[%
    0.45\textwidth %
]%
{\raisebox{\dimexpr.5\ht\largestimage-.5\height}%
{%
    \includegraphics[width=0.45\textwidth]{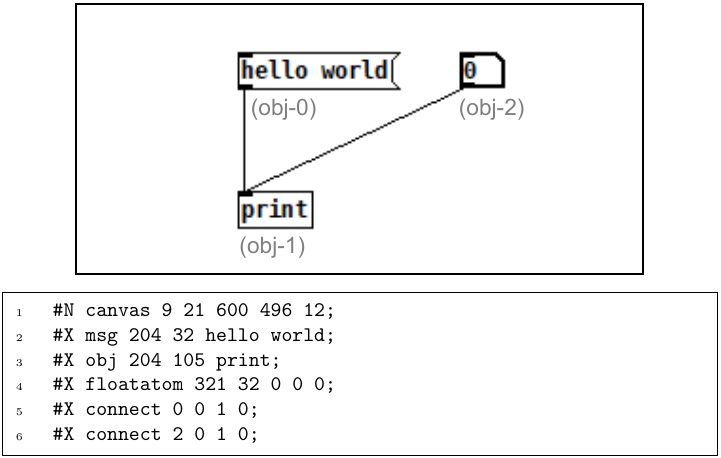}%
}}%
\hspace{0.05\linewidth} %
\subcaptionbox[Short Subcaption]{%
    JSON IR with 3 subtrees corresponding to the nodes.%
    \label{subfig:ir_2}%
}
[%
    0.45\textwidth %
]%
{%
    \usebox{\largestimage}%
}%
\caption[Short Caption]{Pure Data visual code translated into the intermediate format.}
\label{fig:intermediate_format}
\end{figure*}

We use an \textit{intermediate representation} (IR) of the visual code to perform comparisons between different versions of the code. The IR is a tree that consists of subtrees that contain information about nodes, properties of the nodes, and their connections. A general template of an IR can be seen in \Cref{fig:example-IR} where \textit{node\_id} would be the root of a subtree, \textit{connections} would contain information about the node's edges, and \textit{serialized\_contents} would contain information about properties of the node. It should be noted that the \textit{serialized\_contents} can contain more nested subtrees when a node is nested within a node in visual code (e.g.\ Max/MSP contains a \qq{patcher} property which allows the creation of a sub-node patch within a node patch). An example of an IR and its original visual code file format in Pure Data can be seen in \Cref{fig:intermediate_format}.

\begin{figure}[htbp]
\begin{minipage}{0.485\textwidth}
\begin{minted}[frame=single, numbersep=-10pt, linenos, fontsize=\scriptsize]{python}
    {
        'node_id': {
            'connections': [],
            'serialized_contents': {}
        }
    }
\end{minted}
\end{minipage}
\caption{Example JSON representation of the intermediate representation consisting of 1 subtree without details.}
\label{fig:example-IR}
\end{figure}

\begin{figure*}[htbp]
\centering
\savebox{\largestimagedd}{\includegraphics[width=0.45\textwidth]{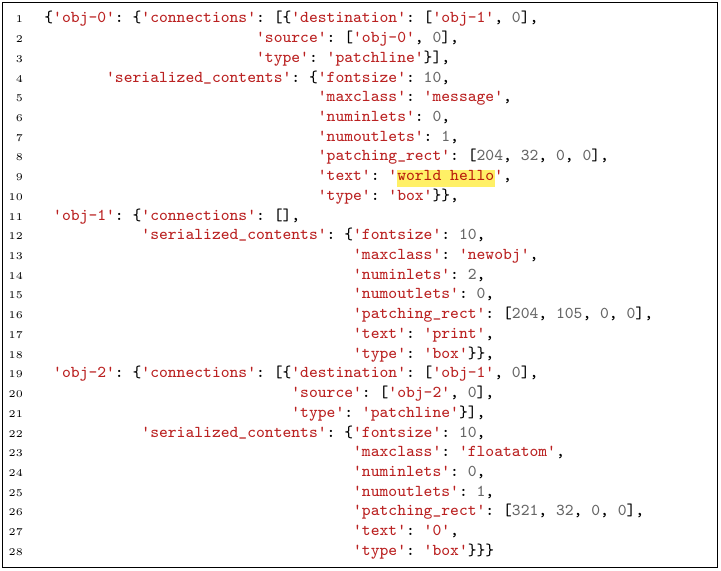}}%
\subcaptionbox[Short Subcaption]{%
    Modified Pure Data visual code IR of \Cref{subfig:ir_2} highlighted with the change from \qq{hello world} to \qq{world hello}.%
    \label{subfig:dd_ir_1}%
}
[%
    0.45\textwidth %
]%
{%
    \usebox{\largestimagedd}%
}%
\hspace{0.05\linewidth} %
\subcaptionbox[Short Subcaption]{%
    DeepDiff output of \Cref{subfig:ir_2} and \Cref{subfig:dd_ir_1} annotated with change-depths.%
    \label{subfig:dd_ir_2}%
}
[%
    0.45\textwidth %
]%
{\raisebox{\dimexpr.5\ht\largestimagedd-.5\height}%
{%
    \includegraphics[width=0.45\textwidth]{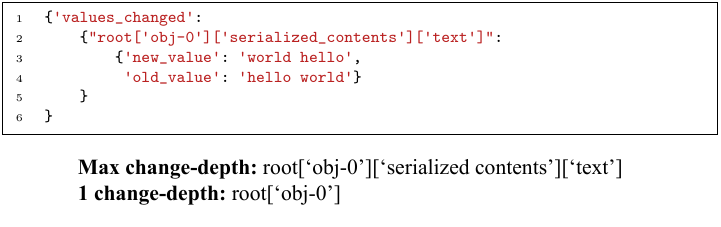}%
}}%

\caption[Short Caption]{Pure Data IR differencing with the Python DeepDiff~\cite{deepdiff} library.}
\label{fig:deepdiff}
\end{figure*}

The core of SZZ-VC is the serialization of visual code into the IR that captures information about nodes and edges. This is because traditional line-by-line code does not capture the semantics of nodes, e.g., the definitions of a node's connections which might be elsewhere in a file. To understand changes in visual code, we can compare subtrees of the IR implemented by DeepDiff~\cite{deepdiff} and represent their changes at different depths. We refer to these granularities of change as \textit{change-depth}. An example of differencing can be seen in \Cref{subfig:dd_ir_2}. In \Cref{subfig:dd_ir_2}, a max change-depth (which is also a change-depth of 3) would be represented as \textit{root[obj-0]['serialized\_contents']['text']} for the changes between \Cref{subfig:ir_2,subfig:dd_ir_1}. The represented max change-depth means that \textit{text} has changed between the \qq{hello world} nodes identified as \textit{obj-0} in \Cref{subfig:ir_2,subfig:dd_ir_1}. A change-depth of 1 would be \textit{root[obj-0]} and mean that there have been changes to at least one element in either \textit{connections} or \textit{serialized\_contents} in the subtree of the \qq{hello world} node identified as \textit{obj-0} in \Cref{subfig:ir_2,subfig:dd_ir_1} without any details.

The changes are represented as 3 types: addition, modification, and deletion. To find a defect-inducing commit, we look for the changes that have changed the modified or deleted changes of a subtree in a defect-fixing commit. For changes that are additions in a defect-fixing commit, changes to that change cannot be found since the change is a new addition to the subtree. However, it is still possible to find an inducing change by going up change-depths (e.g.\ 3 to 2) in order to find a defect-inducing commit at the new change-depth. This is similar to the technique done by \citet{sahal2018identifying} who attempt to find defect-inducing changes by looking for changed code blocks of code additions since the new lines may not have any linked changes yet --- each level of the subtree can be considered as a code block.

\subsection{Limitations}
\label{limit_szz_vc}
The limitations of the first part of SZZ-VC in steps (1) and (2) that identify suitable defect-fixing commits are similar to the SZZ algorithm where defect-fixing commits need to be accurately identified and carefully be considered in terms of number of files and the type of defect being fixed. The heuristics to determine what is a suitable defect-fixing must also be carefully considered. For instance, defect-fixing commits may be false negatives if visual code file extension types are missing in the heuristic (e.g.\ a \qq{.maxhelp} extension is also a patch file in Max/MSP in addition to \qq{.maxpat}).

For the limitations of the second part of SZZ-VC in steps (3) and (4), it should be noted that the serialization into the intermediate format is language dependent. This means that for each language, a new serializer will need to be written. The intermediate format is also not standardized like abstract syntax trees, hence one intermediate format may not be directly compared with another language's intermediate format.

It is important to consider the inclusion and exclusion of visual code features into the intermediate format. For example, positional features of where to place a graphical element may not affect functionality. Thus, implementors should address if position matters and adjust the properties of connections or nodes to address this. For example, the \textit{text} property in Max/MSP is important to identify the type of node, but the \textit{patching\_rect} property can be omitted because refers to the size of the node in the editor and is not important to the program logic.   

Finally, depending on parser implementation, which is often independent of visual programming language implementations, SZZ-VC may not be able to deal with syntax errors in the text representation of visual code. Therefore, in the case where files are manually edited, independent of a visual programming editor, SZZ-VC can fail.

\section{Evaluation in Open Source Code}
\label{eval-open-source}
We explore how SZZ-VC can detect visual code defects and its potential issues in the real-world code of 20 music visual programming defect-fixing changes. Our selection of defect-fixing changes is limited in scope because we wish to manually validate the results of the defect-inducing changes found. We use 10 open source Max/MSP defects and 10 Pure Data defects for evaluating SZZ-VC at 1 change-depth and max change-depth. For each of the defects, we compare SZZ-VC results with textual SZZ~\cite{kim2006automatic} results to demonstrate the feasibility of SZZ-VC.

The following subsections explain the implementation of the algorithms, project/issue selection, manual validation method, and the results of our evaluation. We also discuss the benefits and limitations of SZZ-VC over textual SZZ and vice-versa. 

\subsection{Implementation} 
PyDriller~\cite{PyDriller} is used to get commit and file data for textual SZZ and SZZ-VC. PyDriller is also used to implement a version of textual SZZ~\cite{kim2006automatic}. Our implementation of SZZ-VC uses modified parsers, that were previously used to find code clones by Burlet \textit{et al.}~\cite{burlet2015empirical}, to translate Max/MSP and Pure Data code into an intermediate representation. To perform differencing in SZZ-VC, we use the Python DeepDiff~\cite{deepdiff} library that performs a recursive depth first search to compare contents of Python objects at different depths. Specifically for the intermediate representation, we search for changes in connections and contents between the subtree node objects of different commits. We run SZZ-VC and textual SZZ only on \textit{.maxpat} or \textit{.maxhelp} files for Max/MSP and \textit{.pd} files for Pure Data. Our implementations of SZZ-VC and textual SZZ can be found in our replication package~\cite{reppackage}.

\subsection{Issue/Project Selection}
For finding Pure Data defects, we manually review the externals in the Pure Data package manager \qq{Deken}~\cite{deken} and try to find the relevant project repositories for the externals using Google search. We identify 232 externals as of March 23, 2023, with 94 distinct GitHub project repositories. From the Pure Data GitHub repositories, we look for closed issues in their GitHub issue trackers and find a total of 789 issues as of May 9, 2023. For each of the 789 issues, we check to see if they reference a commit that modifies at least one \textit{.pd} file and get a total of 205 issues. We manually review each of the 205 issues and try to determine if they are referring to defects, if they are extrinsic or intrinsic defects, rate them in terms of fix explainability (1=explainable, 0.5=possible or inferred explanation, 0=unexplained), and extract the defect-fixing commits. From the manual review, we filter down to 24 issues that refer to intrinsic defects, are explainable, and have a fixing commit. We further reduce the number of issues by keeping the commits that have only modified one \textit{.pd} file. In total, we end up with 10 Pure Data defect-fixing commits for evaluation.

To find Max/MSP defects, we use the GitHub search API to search for closed issues that are labelled as \textit{bug} and use the \textit{Max} language. We retrieved 703 issues as of May 2, 2023. For each of the issues, we check to see if they reference a commit that modifies at least one \textit{.maxpat} or \textit{.maxhelp} file and get a total of 98 issues. We manually review each of the 98 issues and try to determine if they are referring to defects, if they are extrinsic or intrinsic defects, rate them in terms of fix explainability (1=explainable, 0.5=possible or inferred explanation, 0=unexplained), and extract the defect-fixing commit. We filter down to 38 issues that refer to intrinsic defects, are explainable, and have a fixing commit. We further reduce the number of issues by keeping the commits that have only modified one \textit{.maxpat} or \textit{.maxhelp} file. In total, we end up with 10 Max/MSP defect-fixing commits for evaluation.

We note that not all defects in the issues are equal, i.e., defects can be \textit{intrinsic} when found in the code, and \textit{extrinsic} when defects are caused by external factors such as changes in dependency API dependencies, changing requirements, or system specific issues~\cite{rodriguez2020bugs, rodriguez2018if}. Since extrinsic defects cannot be found by SZZ due to external influences of the functionality of code, we preclude them from our potential defect-fixing commits.

In total, there remains 4 Pure Data projects with 10 ($6+1+1+1+1$) defects and 8 Max/MSP projects with 10 ($2+2+1+1+1+1+1+1$) defects to evaluate SZZ-VC.  More details of our mining process to gather projects can be found in our replication package~\cite{reppackage}. \Cref{tab:ground-truth} presents the issues used for evaluation including: project name, visual code language, issue description, fixing commit, and the fix we infer from issue and commit data. 

\begin{table*}[tbp]
\centering
\caption{Defect-fixing changes to verify SZZ. (FC = Fixing Commit, LoCC = Lines of Code Changed, NT= Nodes Touched)}
\label{tab:ground-truth}
\begin{tblr}{
  width = \linewidth,
  colspec = {Q[150]Q[88]Q[458]Q[71]Q[135]Q[135]},
  hlines,
  vline{2-6} = {-}{},
  hline{1,22} = {-}{0.08em},
}
\textbf{Project} & \textbf{Language} & \textbf{Issue}                                                                                & {\textbf{FC}} & {\textbf{FC LoCC}} & {\textbf{FC \# of NT}} \\
Gem~\cite{gem}              & Pure Data         & {[}primTri] argument issue                                                                    & ac27153                             & 1                                                                  & 1                                                                 \\
Gem~\cite{gem}              & Pure Data         & Stereo 2 screen
  mode is broken                                                              & 42a2c50                             & 10                                                                 & 5                                                                 \\
Gem~\cite{gem}              & Pure Data         & {[}ortho] has wrong buffer size                                                               & bda2c4b                             & 2                                                                  & 1                                                                 \\
Gem~\cite{gem}              & Pure Data         & Example
  03.movement\_detection.pd has inverted Y-axis logic                                 & 026159e                             & 4                                                                  & 2                                                                 \\
Gem~\cite{gem}              & Pure Data         & gemwin transparency
  doesn't work for the first window creation                              & f6184ca                             & 4                                                                  & 11                                                                \\
Gem~\cite{gem}              & Pure Data         & clear buffer in
  single buffer mode                                                          & 88e0e34                             & 18                                                                 & 2                                                                 \\
Cyclone~\cite{cyclone}          & Pure Data         & {[}seq] object doesn't recognize "play" message                                               & d9827cc                             & 66                                                                 & 37                                                                \\
Cyclone~\cite{cyclone}          & Pure Data         & accum-help.pd and
  number boxes                                                              & 93e8db8                             & 109                                                                & 36                                                                \\
Else~\cite{else}             & Pure Data         & Index error in [rand.list]                                                                    & d91382b                             & 18                                                                 & 8                                                                 \\
Deken~\cite{deken-plugin}            & Pure Data         & README.deken.pd
  broken                                                                      & 55713aa                             & 2                                                                  & 2                                                                 \\
264 Tools~\cite{264tools}        & Max/MSP~ ~        & Looping at negative
  playback rates doesn't work in 264.sfplay\textasciitilde{}              & 2634806                             & 2,545                                                              & 5                                                                 \\
264 Tools~\cite{264tools}        & Max/MSP~ ~        & Fix mislabelled
  third inlet to 264.loop\textasciitilde{}                                    & 9bd28f6                             & 13                                                                 & 1                                                                 \\
VideoAnalysis~\cite{videoanalysis-1}    & Max/MSP~ ~        & "Noise
  Reduction" toggle should decide noise reduction, not the actual n.r.
  value         & 07b5428                             & 101                                                                & 10                                                                \\
Sonorium~\cite{sonorium}         & Max/MSP~ ~        & Bug — selecting presets                                                                       & 1b423d4                             & 215                                                                & 20                                                                \\
odot~\cite{odot}             & Max/MSP~ ~        & o.listenumerate
  helpfile produces errors when a message has no data                         & 4d4192f                             & 102                                                                & 6                                                                 \\
Cut Glove~\cite{cut-glove}        & Max/MSP~ ~        & Can't MIDI learn
  record/play (due to them being ubuttons)                                   & 69459fe                             & 3,174                                                              & 10                                                                \\
FluCoMa~\cite{flucoma}         & Max/MSP~ ~        & fluid.bufselectevery\textasciitilde{}
  help file requests @numderivs 3, which isn't possible & a78a5ba                             & 271                                                                & 1                                                                 \\
FluCoMa~\cite{flucoma}          & Max/MSP~ ~        & fluid.grid\textasciitilde{} help
  file example not working                                   & ca95828                             & 32                                                                 & 1                                                                 \\
VideoAnalysis*~\cite{videoanalysis-2}    & Max/MSP~ ~        & Flickering videos                                                                             & c701997                             & 15,869                                                             & 2                                                                 \\
FrameLib~\cite{framelib}         & Max/MSP~ ~        & Max tutorial 8 typo                                                                           & d00b4f4                             & 191                                                                & 7                                                                 
\end{tblr}
\end{table*}

\subsection{Validation Method}
\label{subsec:val-method}
To see how SZZ-VC compares against textual SZZ which performs text based differencing, we manually review the defect-inducing changes that are detected by each method. Textual SZZ is possible because the visual code is saved in a textual format.
The changes are reviewed by an author of this paper that has proficient experience in parsing Max/MSP and Pure Data source code by implementing parsers for them as well as a basic understanding of the Max/MSP and Pure Data editors. During the review process, 10 minutes is given to each defect-inducing commit to determine if the discovered commit is a \textit{true positive} or \textit{false positive}. If a decision cannot be made within 10 minutes, then the commit is marked as \textit{unknown}. Also, during the review process, comments are recorded about why the decision is made for each commit.

To understand each of the defect-inducing commits in textual SZZ, we look at the changes to the deleted changed lines of a defect-fixed commit and try to find context around the lines of changed code (e.g.\ which node a modified line corresponds to). While for SZZ-VC we look at the changes to the modified or deleted parts of the IR in the defect-fixing commit. Observations and concerns about the commit contents are also noted down while reviewing each commit.

\textit{True positives} (TP) refers to the detected defect-inducing commits that we analyze and suspect to be causing a defect, while \textit{false positives} (FP) refers to the commits that are detected but we suspect not to cause a defect. To compare our TP and FP across algorithms and commits, we calculate precision (i.e.\ $\frac{\text{TP}}{\text{TP}+\text{FP}}$). A precision closer to 1 means that we think SZZ is very accurate in identifying defect-inducing commits. We can evaluate in only these terms of measures since we do not have the necessary domain expertise to provide a ground truth for the valid defect-inducing commits of a defect-fixing commit to identify the false negatives needed for recall and the F1 score.

\subsection{Results}
The results of our evaluation in terms of \textit{true positives} (TP), \textit{false positives} (FP), \textit{unknown} (U), and \textit{precision} (Pr) for the defect-inducing commits of SZZ-VC (1 change-depth), SZZ-VC (max change-depth), and textual SZZ are presented in \Cref{tab:results}. The best values for each row are underlined.

In \Cref{tab:results}, we see that SZZ-VC (max change-depth) performs better than the other variations in Max/MSP code with an average precision of $0.93$. In contrast, it appears that textual SZZ is slightly better than the other algorithms for Pure Data code with an average precision of 0.55. The slightly better performance of textual SZZ in Pure Data code might be because textual SZZ does not explicitly identify their nodes, meaning it can track changes independent of modified node identifiers. Overall, we conclude that SZZ-VC achieves better performance than textual SZZ on Max/MSP code. For Pure Data code, it appears that textual SZZ achieves a slight advantage over SZZ-VC. We discuss why some results may be better than others in \Cref{subsec:discuss}.

\begin{table*}[tbp]
\centering
\caption{Performance of SZZ-VC variations and textual SZZ per defect-fixing commit from manual validation (TDIC = Total Defect-Inducing Change)}
\begin{tblr}{
  width = \linewidth,
  colspec = {Q[96]Q[48]Q[48]Q[37]Q[71]Q[58]Q[54]Q[54]Q[40]Q[81]Q[54]Q[54]Q[54]Q[40]Q[79]Q[52]},
  cell{1}{2} = {c=5}{0.262\linewidth},
  cell{1}{7} = {c=5}{0.283\linewidth},
  cell{1}{12} = {c=5}{0.278\linewidth},
  cell{13}{1} = {c=5}{0.3\linewidth,r},
  cell{13}{7} = {c=4}{0.228\linewidth,r},
  cell{13}{12} = {c=4}{0.226\linewidth,r},
  cell{24}{1} = {c=5}{0.3\linewidth,r},
  cell{24}{7} = {c=4}{0.228\linewidth,r},
  cell{24}{12} = {c=4}{0.226\linewidth,r},
  vline{2,5-7,10-12,15-16} = {2-12,14-23}{},
  vline{7,12} = {13,24}{},
  hline{1,25} = {-}{0.08em},
  hline{2-3,4-24} = {-}{},
  hline{4} = {1-14,16}{},
}
                             & \textbf{SZZ-VC (1 change-depth) } &    &   &      &             & \textbf{SZZ-VC (Max change-depth) } &    &   &      &              & \textbf{Textual SZZ}  &    &   &      &              \\
\textbf{Commit}              & TP                                & FP & U & TDIC & Pr          & TP                                  & FP & U & TDIC & Pr           & TP                    & FP & U & TDIC & Pr           \\
ac27153                      & 1                                 & 0  & 0 & 1    & \uline{1.0} & 1                                   & 0  & 0 & 1    & \uline{1.0}  & 1                     & 0  & 0 & 1    & \uline{1.0}  \\
42a2c50                      & 2                                 & 0  & 0 & 2    & \uline{1.0} & 2                                   & 0  & 1 & 3    & \uline{1.0}  & 2                     & 0  & 0 & 2    & \uline{1.0}  \\
bda2c4b                      & 1                                 & 0  & 0 & 1    & \uline{1.0} & 1                                   & 0  & 0 & 1    & \uline{1.0}  & 1                     & 0  & 0 & 1    & \uline{1.0}  \\
026159e                      & 0                                 & 1  & 0 & 1    & 0           & 0                                   & 1  & 0 & 1    & 0            & 0                     & 1  & 0 & 1    & 0            \\
f6184ca                      & 0                                 & 1  & 1 & 2    & 0           & 0                                   & 1  & 1 & 2    & 0            & 0                     & 0  & 1 & 1    & 0            \\
88e0e34                      & 1                                 & 0  & 0 & 1    & \uline{1.0} & 1                                   & 0  & 0 & 1    & \uline{1.0}  & 1                     & 0  & 0 & 1    & \uline{1.0}  \\
d9827cc                      & 0                                 & 1  & 7 & 8    & 0           & 0                                   & 0  & 9 & 9    & 0            & 0                     & 0  & 5 & 5    & 0            \\
93e8db8                      & 0                                 & 0  & 4 & 4    & 0           & 0                                   & 0  & 6 & 6    & 0            & 0                     & 0  & 3 & 3    & 0            \\
d91382b                      & 0                                 & 3  & 0 & 3    & 0           & 0                                   & 3  & 0 & 3    & 0            & 1                     & 1  & 0 & 2    & \uline{0.5}  \\
55713aa                      & 0                                 & 1  & 0 & 1    & 0           & 0                                   & 1  & 0 & 1    & 0            & 1                     & 0  & 0 & 1    & \uline{1.0}  \\
Avg Pure Data (PD) Precision &                                   &    &   &      & 0.4         & Avg PD Precision                    &    &   &      & 0.40         & Avg PD Precision      &    &   &      & \uline{0.55} \\
2634806                      & 2                                 & 0  & 0 & 2    & \uline{1.0} & 4                                   & 0  & 1 & 5    & \uline{1.0}  & 2                     & 2  & 2 & 6    & 0.5          \\
07b5428                      & 0                                 & 0  & 2 & 2    & 0           & 1                                   & 0  & 0 & 1    & \uline{1.0}  & 2                     & 0  & 0 & 2    & \uline{1.0}  \\
1b423d4                      & 1                                 & 0  & 1 & 1    & \uline{1.0} & 1                                   & 0  & 1 & 2    & \uline{1.0}  & 2                     & 2  & 0 & 4    & \uline{0.5}  \\
4d4192f                      & 1                                 & 0  & 1 & 1    & \uline{1.0} & 2                                   & 0  & 1 & 3    & \uline{1.0}  & 2                     & 0  & 1 & 3    & \uline{1.0}  \\
69459fe                      & 1                                 & 0  & 1 & 2    & \uline{1.0} & 1                                   & 0  & 1 & 2    & \uline{1.0}  & 0                     & 0  & 7 & 7    & 0            \\
9bd28f6                      & 1                                 & 0  & 0 & 1    & \uline{1.0} & 1                                   & 0  & 0 & 1    & \uline{1.0}  & 1                     & 0  & 0 & 1    & \uline{1.0}  \\
a78a5ba                      & 2                                 & 0  & 0 & 2    & \uline{1.0} & 2                                   & 0  & 0 & 2    & \uline{1.0}  & 0                     & 3  & 1 & 4    & 0            \\
c701997                      & 1                                 & 1  & 0 & 2    & 0.5         & 2                                   & 0  & 0 & 2    & \uline{1.0}  & 0                     & 0  & 5 & 5    & 0            \\
ca95828                      & 0                                 & 1  & 0 & 1    & 0           & 1                                   & 0  & 0 & 1    & \uline{1.0}  & 0                     & 2  & 0 & 2    & 0            \\
d00b4f4                      & 0                                 & 3  & 0 & 3    & 0           & 1                                   & 2  & 0 & 3    & \uline{0.33} & 0                     & 7  & 1 & 8    & 0            \\
Avg Max/MSP Precision        &                                   &    &   &      & 0.65        & Avg Max/MSP Precision               &    &   &      & \uline{0.93} & Avg Max/MSP Precision &    &   &      & 0.4          
\end{tblr}
\label{tab:results}
\end{table*}

\subsection{Discussion}
\label{subsec:discuss}
Using SZZ-VC comes with benefits and limitations over textual SZZ. We discuss why textual comparison is hard, why change-depths matter, how syntactic and irrelevant changes can impact outcomes of SZZ.

\subsubsection{Textual Comparison is Hard}
While manually validating the defect-inducing commits, notes were made several times about textual code being difficult to understand and compare leading to many unknowns. For example, commit \textit{c701997} changes 15,869 lines of code and SZZ-VC is able to identify 2 nodes were touched (added/modified/deleted). Changes of that size are not possible to trace manually within 10 minutes. Therefore, the comparisons at the semantic level by SZZ-VC is an advantage. We suggest that a visual differencing tools need to be implemented in order to better comprehend visual code changes. 

\subsubsection{SZZ-VC Change-Depths}
When comparing SZZ-VC (1 change-depth) and SZZ-VC (max change-depth), we can see in \Cref{tab:results} that not all results in terms of precision are equal. This is because of the different granularities of change, and the ability of SZZ-VC to detect only changes that have occurred at a certain depth in the past. It should be noted that subtrees of the IR can be exceptionally deep when it contains nested subtrees. At max change-depth, exceptionally specific changes are identified within a subtree that could be irrelevant to a defect-fix. With 1 change-depth, any prior changes to a subtree are used to detect inducing-changes which may not be specific enough. There is a trade-off in terms of precision by setting different depths of comparison.

\subsubsection{Syntactic Changes}
In the Max language, rearranging objects in the visual interface often rearranges the order of code in the textual source code. This makes tracking the line-changes of code difficult and as a result, textual diffs from version control often works poorly in Max patches. We noticed this in commit \textit{69459fe}, making it difficult to understand and comprehend what has exactly changed. Our approach of translating the textual source code into an intermediate representation mitigates the issue of tracking line-changes as we perform subtree differencing that outputs the differences in nodes. 

\subsubsection{Ignoring Irrelevant Changes}
Irrelevant changes such as the position of nodes in visual code can often introduce false positives when positional information is not critical to the functionality of a visual programming language. One such case is commit \textit{1b423d4} where we attribute a FP change to a non-functional line change. Since positional information is embedded in the textual representation of visual code, simple rearrangements of nodes in visual code can indicate that a change is defect-inducing when it does not affect any functionality. The translation of changes into an intermediate representation in SZZ-VC allows for the understanding of context in the textual source code file to capture only important changes as opposed to line based differencing in textual SZZ. Therefore, we suggest implementors of SZZ-VC to decide which properties are relevant to their scenario. 

\subsubsection{Persistent IDs}
In visual programming languages such as Pure Data, the IDs of a node are identified by the order that it is defined in the textual code, as opposed to Max/MSP which explicitly defines IDs for each node and makes them persistent across changes. Persistent IDs make it possible to track for changes even if lines are rearranged in Max/MSP. Since Pure Data does not define any IDs and pseudo-IDs are created based on the ordinal appearance of node definitions in a file, it is unable to track changes if additional lines are added to define a new node in-between the lines of existing textual code. We suspect this is the main reason for such a low precision for SZZ-VC in \Cref{tab:results}. To better address this in Pure Data, better mechanisms are needed to track changes to nodes in Pure Data that are changed such as employing similarity measures to map nodes from commit-to-commit.

\section{Application in Industry}
\begin{figure*}[htbp]
\centering
\savebox{\largestimageind}{\includegraphics[width=0.45\textwidth]{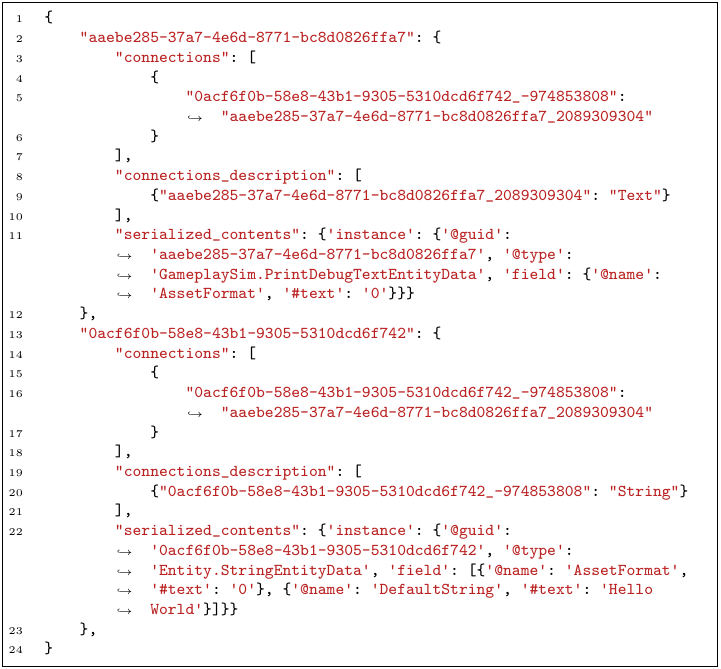}}%
\subcaptionbox[Short Subcaption]{%
    Visual code from internal visual scripting engine annotated with the IR root subtree id's%
    \label{subfig:industry1}%
}
[%
    0.45\textwidth %
]%
{\raisebox{\dimexpr.5\ht\largestimageind-.5\height}%
{%
    \includegraphics[width=0.45\textwidth]{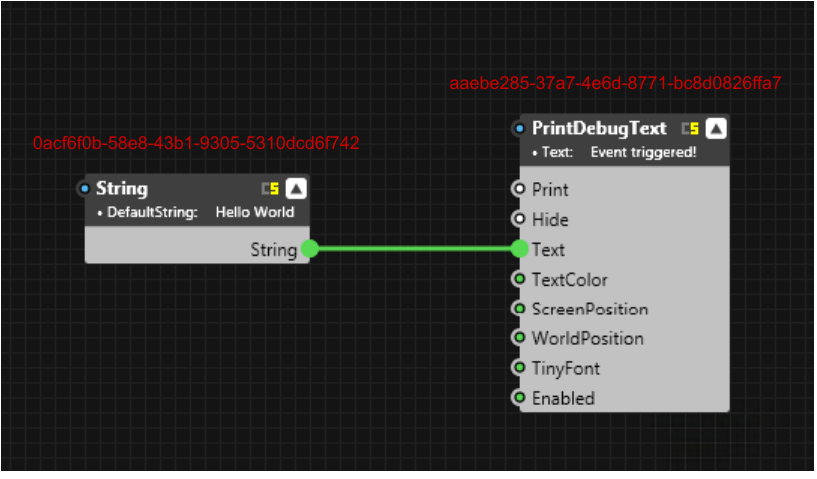}%
}}%
\hspace{0.05\linewidth} %
\subcaptionbox[Short Subcaption]{%
    JSON IR with 2 subtrees identified by GUID%
    \label{subfig:industry2}%
}
[%
    0.45\textwidth %
]%
{%
    \usebox{\largestimageind}%
}%
\caption[Short Caption]{Visual code of internal visual scripting engine translated into the intermediate representation.}
\label{fig:intermediate_format_industry}
\end{figure*}

SZZ-VC has been applied at \textit{EA} (Electronic Arts) to detect defect-inducing changes for a AAA video game that was primarily developed using visual code. The purpose for finding defect-inducing changes in the AAA video game is to gather defect-inducing targets for building defect prediction models~\cite{senchenko2022supernova} that help improve workflows and reduce testing time. The visual code is developed using an internal data-driven visual scripting engine that gives game designers the freedom to design any type of gameplay experience with limited programming knowledge. The visual scripting is similar to Pure Data and Max/MSP since programming is done via nodes and edges. An example of the visual code can be seen in \Cref{fig:intermediate_format_industry}.

We evaluate SZZ-VC against textual SZZ similar to \Cref{subsec:val-method}. This evaluation is different from the open source evaluation because we manually evaluate a substantial amount of issues and commits for a single repository. This enables us to test SZZ-VC at scale. In contrast, we were only able to find a small amount of well-labelled defect-fixing commits in open source.

We found 3,044 defect-fixing commits related to changes that contain visual programming files in the issue tracker as of March 20, 2023. Notably, the issue reporters were trained on how to succinctly report issues ensuring high quality descriptions and linking of artifacts which helps with manual verification. Most of the open source projects did not have such an advantage. To evaluate the SZZ implementations, we randomly sample 30 defect-fixing commits. The 30 commit sample size is determined using Cochran’s formula with a 90\% confidence level, maximum variability of 0.5, and a precision of 15\%. The sample size indicates that the conclusion about SZZ-VC is within 15\% of the real population value 90\% of the time.

We present our results for the 30 commits in terms of \emph{total true positives} (TTP), \emph{total false positives} (TFP), \emph{total unknown} (TU), \emph{total defect-inducing changes} (TDIC) and \emph{average precision} (AP) for SZZ-VC (depth=1), SZZ-VC (max depth), and textual SZZ in \Cref{tab:ea-results}. We present the precision in terms of averages to summarize the performance of defect-inducing commit detection across the 30 chosen commits. It should be noted that textual SZZ could not find defect-inducing commits for only 22 defect-fixing commits, hence we assign a precision of 0 for 8 of the textual SZZ results.

In the \Cref{tab:ea-results}, SZZ-VC has better performance than textual SZZ which is similar to our open source evaluation. Hence, we can conclude that SZZ-VC is feasible for detecting defects in visual code for 3 different visual programming languages. This manual validation of defect-inducing changes also demonstrates that SZZ-VC is effective in detecting defect-inducing changes at scale in comparison with textual SZZ. Thus, SZZ-VC can be used to find defect-inducing labels to build visual code defect prediction models.

\begin{table}[tbp]
\centering
\caption{Average Performance of SZZ-VC variations and textual SZZ in industry-made AAA video game.}
\begin{tblr}{
  hlines,
  vline{2,5-6} = {-}{},
}
\textbf{Method}           & \textbf{TTP} & \textbf{TFP} & \textbf{TU} & \textbf{TDIC} & \textbf{AP} \\
SZZ-VC (1 change-depth)   &  41                 &  5                 &  8                & 54                  & 0.79            \\
SZZ-VC (max change-depth) &  57                 &  12                 &  2                & 71                   & \uline{0.87}            \\
Textual SZZ               &  27                 &   12                &   2               & 41                   & 0.62            
\end{tblr}
\label{tab:ea-results}
\end{table}

\section{Threats to Validity}
Internal validity concerns our ability to identify the bug reports that match defect-fixing commits, the code changes that are defect-fixing, and the code changes that are defect-inducing in the specific project defects we selected to test SZZ-VC. We highlight the limitations to these in \Cref{og-szz} and \Cref{limit_szz_vc}. To these extents, we may not have identified all potential defect-fixing or defect-inducing commits in our SZZ evaluations.

Construct validity refers to the validity of assessment relevant to our conclusions about the accuracy of the SZZ-VC in identifying visual code defects. We rely on the paper authors' conclusions about the accuracy of defect-fixing and defect-inducing changes instead of seeking validation from the original contributors to the changes. However, during validation efforts were made to verify a change is correct by referring to explanation of fixes in bug reports and commit messages. We make our open-source data publicly available in a replication package~\cite{reppackage} so that our method can be verified by anybody who wishes to do so.

External validity considers the extent to which the SZZ-VC can be applied to other projects that use visual programming languages, beyond those we selected for our testing. This validity is limited by the lack of open source visual code defects to evaluate. We have demonstrated the applicability of our algorithm for 3 different visual programming languages. 

\section{Conclusion}
In this paper, we present the SZZ Visual Code (SZZ-VC) approach to identify defects in visual code. SZZ-VC compares the changes of nodes and edges in visual code rather than implementing line-by-line comparisons. We evaluate SZZ-VC for 20 music visual programming defects in 12 open source projects, and 30 defects in an industrial AAA video game project demonstrating its practicality in detecting defects in visual programming. Our evaluations show that SZZ-VC performs better in cases when nodes are explicitly defined in visual code and suggests that it has potential to help identify the root cause of visual code defects where textual SZZ cannot. SZZ-VC and its IR make it easier to determine what the actual defect was. Our approach helps identify defect-inducing changes which is essential to improve code quality, evaluate testing methods, and support machine learning software defect prediction. We make our results of our open source analysis available in a replication package~\cite{reppackage}.

\bibliographystyle{IEEEtranN-ital}
\bibliography{main}

\end{document}